\documentclass[onecolumn,showkeys,preprintnumbers,aps,a4paper,amssymb,prd,superscriptaddress,nofootinbib]{revtex4-2}
\usepackage{comment}
\usepackage{graphicx}
\usepackage{epsf}
\usepackage{bm}
\usepackage{amsmath}
\usepackage{amsfonts}
\usepackage{amssymb}
\usepackage{epstopdf}
\usepackage{color}
\usepackage[dvipsnames]{xcolor}
\usepackage{verbatim}
\usepackage{multirow}
\usepackage{soul}
\usepackage{physics}
\usepackage{bm}

\usepackage[width=0.00cm, height=0.00cm, left=1.50cm, right=1.50cm, top=2.00cm, bottom=2.00cm]{geometry}
\usepackage{microtype}
\usepackage{lmodern}

\usepackage[colorlinks = true,
linkcolor = teal,
urlcolor  = teal,
citecolor = teal,
anchorcolor = blue]{hyperref}
\usepackage[capitalize]{cleveref}
\usepackage[normalem]{ulem}
\usepackage{enumitem}
\usepackage{booktabs}

\usepackage{lipsum}

\makeatletter\let\expandableinput\@@input\makeatother

\begin{document}
\begin{center}
		\vspace{0.4cm} {\large{\bf Observational Constraints and Geometric Diagnostics of Barboza-Alcaniz and Logarithmic Dark Energy Parametrizations}
} \\
		\vspace{0.4cm}
		\normalsize{Archana Dixit$^1$, Saurabh Verma$^2$, Anirudh Pradhan$^3$, M. S. Barak$^4$ }\\
		\vspace{5mm}
		\normalsize{$^{1}$ Department of Mathematics, Gurugram University, Gurugram, Harayana, India\\
		\normalsize{$^{2,4 }$ Department of Mathematics, Indira Gandhi University, Meerpur, Haryana 122502, India }\\ 
	
		\normalsize{$^{3 }$ Centre for Cosmology, Astrophysics and Space Science (CCASS), GLA University, Mathura-281406, Uttar Pradesh, India}\\ 

		\vspace{2mm}
		$^1$Email address: archana.ibs.maths@gmail.com\\
		$^2$Email address: saurabh.math.rs@igu.ac.in\\
		$^3$Email address: pradhan.anirudh@gmail.com\\
        $^4$Email address: ms$_{-}$barak@igu.ac.in\\}
\end{center}

 
\maketitle
%
{\bf Abstract}:
This study investigates and compares two prominent two-dimensional dark energy (DE) parameterizations: Barboza-Alcaniz (BA) and Logarithmic forms by comparing them with a comprehensive set of observational data comprising Type Ia Supernovae (SNe Ia) from the Pantheon compilation, Baryon Acoustic Oscillations (DESI BAO), and Cosmic Chronometers (CC). The primary objective was to explore the constraining power and cosmological implications of each parameterization in light of the current data. After formulating the theoretical framework and background equations governing cosmic expansion, we employ Markov Chain Monte Carlo (MCMC) techniques using the emcee Python package to constrain the free parameters of each model. The best-fit values for parameters $\omega_0$, $\omega_a$, and $H_0$ were extracted for each model using individual and combined datasets. The results include confidence contours at the levels $1\sigma$ and $2\sigma$. Our findings demonstrate that both parameterizations are consistent with observational data, with logarithmic parameterization showing slightly better constraints in terms of parameter evolution. Furthermore, we employed a statefinder diagnostic to analyze the geometric behavior of the models, providing an effective distinction between the two DE scenarios. This study contributes to a deeper understanding of DE evolution and its constraints in light of current cosmological data.

\smallskip 


\section{Introduction}

The discovery that the Universe is undergoing accelerated expansion was first made in the late 1990s through extensive observations of high-redshift  Type Ia supernovae (SNe Ia) \cite{ref1,ref2}. Subsequent developments in observational cosmology have further confirmed this acceleration phase. To explain this phenomenon, cosmologists have proposed various models, that can be broadly classified into two main categories based on their underlying theoretical assumptions. The first group state that faster expansion is caused by an unknown part with negative pressure, which is often called dark energy (DE). The second category explains the observed acceleration by modifying the standard theory of gravity on extragalactic scales \cite{ref3,ref4,ref5}. Within the framework of general relativity (GR), the cosmological constant ($\Lambda$) is considered the simplest and most natural candidate for dark energy (DE) occupying approximately $70 \%$ of the energy budget of the universe \cite{ref6,ref7,ref8}. By including the cosmological constant ($\Lambda$) alongside the cold dark matter (CDM), one can readily explain the universe’s current phase of accelerated expansion. Despite its strong observational success, the $\Lambda$CDM model faces several theoretical challenges \cite{ref9,ref10,ref11}.\\

Furthermore, the $\Lambda$CDM model faces several notable observational tensions regarding the estimation of the key cosmological parameters. First, the Lyman-$\alpha$ forest measurement of Baryon Acoustic Oscillations \cite{ref12} indicated a lower value of the matter density parameter compared to the value inferred from Cosmic Microwave Background (CMB) data. Second, there is a discrepancy between large-scale structural observations \cite{ref13}. Additionally, statistically significant tension exists between the value of the Hubble constant ($H_0$) derived from the classical distance ladder method and that measured by Planck CMB observations \cite{ref14}. Quantitatively speaking, the $\Lambda$CDM cosmology deduced from Planck CMB data predict $H_0 = 67.4\pm0.5$ km/s/Mpc \cite{ref15}, while from the Cepheid-calibrated SnIa \cite{ref16} we have $H_0=74.03\pm 1.42$ km/s/Mpc, while $\Lambda$CDM provides a straightforward and efficient explanation, which has a number of theoretical issues, including the coincidence and fine-tuning issues \cite{refa,refb}.  These problems have spurred research on other theories, such as General Relativity modifications and scenarios with a dynamic type of dark energy. Several studies employing model-independent cosmographic methods reported in the literature suggest that the significant tension observed between the predictions of the $\Lambda$CDM model and those derived from model-independent analyses supports the idea that it is worthwhile to search for viable alternatives to the $\Lambda$CDM paradigm \cite{ref17,ref18,ref19,ref20,ref21}.\\

Several dark energy models have been proposed as alternatives to $\Lambda$CDM. The simplest and most natural candidate is the cosmological constant \cite{New1}; however, other scenarios such as quintessence \cite{New2}, phantom energy \cite{New3}, and various dynamical parameterizations have also been considered. Widely used parameterizations include the Chevallier–Polarski–Linder (CPL) form \cite{New4,New5}, Barboza–Alcaniz (BA)  \cite{New6}, and Logarithmic model \cite{New7}, each aiming to capture possible deviations from the cosmological constant at different redshift ranges. Moving beyond the $\Lambda$CDM framework, researchers have proposed various DE models, including the $w$CDM model, which assumes DE as a perfect fluid with a constant state parameter $w$, differing from $-1$ other dynamic DE scenarios in the literature \cite{ref22,ref23,ref24,ref25,ref26,ref27,ref28}. Furthermore one method for examining the dynamical dark energy is to use a redshift-dependent parametrized equation of state (EoS) $w(z)$.  The literature \cite{ref31,ref32,ref33,ref34,ref35,ref36,ref37} has suggested a number of parameterizations for $w(z)$ because the characteristics of dark energy are still mostly unknown.\\

For a large number of scalar ﬁeld models with a range of potentialities, testing all individual models is  diﬃcult. Alternatively, we often use a parameterization of the DE evolution that broadly describes a large number of DE models in the scalar region. When parameterizing $(w=\frac{p}{\rho})$, the DE equation of state is the most common practice. Therefore, to evaluate the dynamic parts of the model, we plotted the EoS parameter $w$ as an element of redshift and the literature \cite{ref38,ref39,ref40,ref41,ref42,ref43,ref44,ref45,ref46,ref47,ref48,ref49} also contains a large number of parameterizations for $w$. The conduct of the EoS parameter of the model has been contrasted with that of some notable EoS parameterizations such as CPL \cite{ref50,ref51,ref52,ref53}, Barboza-Alcaniz(BA)\cite{ref54} and Logarithmic \cite{ref55} for our measurements of the evidence together with the regular $\Lambda$CDM and the constant DE equation of the state model. Constraints on the equation of state (EoS) parameters or parameters have been extensively investigated using various cosmological observables such as the weak lensing power spectrum, baryon acoustic oscillations (BAO),  cosmic microwave background (CMB), and SNe Ia. Leveraging these datasets is essential to improve our understanding of dark energy. At present, measurements derived from supernovae, DESI BAO, and Cosmic Chronometers (CC) provide significant constraints on  EoS values. The standard $\Lambda$CDM model is often expressed through a two-parameter form $(w_0, w_a)$, where $w_0$ represents the current EoS value, and $w_a$ captures its time evolution, which may differ from observational trends.\\

The originality of the present work lies not only in updating observational constraints, but also in providing a comprehensive comparative analysis of two widely used bidimensional dark energy parameterizations using the latest DESI BAO data combined with Pantheon and cosmic chronometers. In particular, the inclusion of statefinder diagnostics allows us to extract deeper physical insights into the future evolution of the universe, distinguishing between quintessence-like and phantom-like behaviors. Furthermore, we demonstrate that these phenomenological models can partially alleviate the $H_0$ tension, thereby offering important clues about late-time cosmic dynamics beyond the $\Lambda$CDM paradigm. We emphasize that the Barboza–Alcaniz and logarithmic parametrizations themselves are not newly proposed in this work; rather, the novelty lies in their unified, data-driven comparison and the extraction of physical implications using modern observational datasets and geometric diagnostics.\\

In this study, we investigated several well-established dark energy parameterizations. First, we analyzed the Chevallier–Polarski–Linder (CPL) model, the limitations of which at high redshift $z$ have been discussed in \cite{ref38,ref39,ref40}. To address these issues, the authors proposed an alternative parameterization that retains the same definitions for $w_0$ and $w_a$. Additionally, we examine the Barboza–Alcaniz (BA) parameterization and logarithmic parameterization. These models describe the dark energy component across both low and high redshift regions, with rapid evolution at low $z$, where the CPL model cannot be seamlessly generalized. The purpose of this study is to evaluate two bidimensional dark energy parameterizations using the latest observational data, including SNe Ia, DESI BAO, and CC observations, and to assess which model provides better constraints. This study integrates these perspectives, focusing on the widely used two-parameter models: the Barboza–Alcaniz (BA) and logarithmic parameterizations. The remainder of this paper is organized follows: Section II outlines the theoretical background and introduces the dark energy parameterizations under study. Section III details the observational datasets and describes the methodology used to constrain the models.  Section IV presents the results on how well these models fit the data. Section V looks at how the deceleration and EoS parameters change with different data sets. Section VI presents a state-finder analysis using these datasets. Finally, Section VII concludes the paper with a discussion of the results and directions for future research.

\section{Background equations and parametrization}\label{sec2}

We stress that the current study is phenomenological in character and uses well-motivated parametrizations of the equation of state to constrain dark energy dynamics. We employ an effective fluid description, which is commonly used in empirical cosmology to capture deviations from the $\Lambda$CDM scenario in a model-independent manner, instead of deriving the models from a particular fundamental action. Without committing to a particular microphysical origin, such parametrizations can successfully simulate a wide variety of underlying theories, such as scalar-field dark energy and modified gravity models. A thorough action-based formulation is outside the purview of this research because the main goal of this work is to examine the observational viability of the Barboza–Alcaniz and logarithmic parametrizations using available data.\\

In a Universe permeated by a fluid described by the effective equation of state parameter $w_{\rm eff} = p_{\rm eff} / \rho_{\rm eff}$, the Einstein field equations (using relativistic units where $8\pi G = c = 1$) take the following form within the framework of a spatially flat and homogeneous Robertson–Walker (RW) spacetime:

\begin{equation}\label{eq1}
3H^2=\rho_{\rm eff},
\end{equation}
\begin{equation}\label{eq2}
2\dot{H}+3H^2=- p_{\rm eff}.
\end{equation}
Here,  $H=\dot{a}/a$ denotes the Hubble parameter, whenever $p_{\rm eff}$ and $\rho_{\rm eff}$ represent the effective pressure and energy density of the cosmic fluid, respectively.

\subsection{Barboza-Alcaniz(BA) Parametrization}
We consider BA parametrization for the EoS parameter $w_{\rm eff}$ of the effective fluid in terms of time $z$, that is,
\begin{equation}\label{eq3}
w_{\rm eff}=w_0+w_a z\left(\frac{1+z}{1+z^2}\right),
\end{equation}
where $w_{\rm eff}$ is the effective equation of state (EoS) parameter of dark energy. It characterizes the ratio of pressure $p_{\rm eff}$ to energy density $\rho_{\rm eff}$ of the dark energy component, that is, $w_{\rm eff}=p_{\rm eff}/\rho_{\rm eff}$; $w_0$ is the present-day value of the dark energy EoS parameter, corresponding to $z=0$. It describes the behavior of dark energy in the current universe. $w_a$ is a parameter that quantifies the degree of evolution in the dark energy EoS with redshift and $z$ is the cosmological redshift, which serves as a proxy for time in cosmological studies. A higher value of $z$ corresponds to earlier times in the universe’s history.\\
Now using \eqref{eq1} and \eqref{eq2} into the relation $w_{\rm eff}=p_{\rm eff}/\rho_{\rm eff} $ and comparing it with \eqref{eq3}, we obtain
\begin{equation}\label{eq4}
-\frac{2\dot{H}+3H^2}{3H^2} = \omega_{0}+\omega_{a} z\left(\frac{1+z}{1+z^2}\right),
\end{equation}
\begin{equation}\label{eq5}
-\frac{2\dot{H}}{3H^2}-1 = w_0+w_a z\left(\frac{1+z}{1+z^2}\right)
\end{equation}
As is well known, the time derivative of the Hubble parameter $\dot{H}$ can be expressed in terms of the redshift $z$ as
\begin{equation}\label{eq6}
\dot{H} = -(1+z) \frac{dH}{dz} H
\end{equation}
By substituting \eqref{eq6} into \eqref{eq5}, we obtain the following relation. 
\begin{equation}\label{eq7}
\frac{2}{3H}(1+z)\frac{dH}{dz}=w_0+w_a z\left(\frac{1+z}{1+z^2}\right)+1
\end{equation}
By performing integration on both sides of \eqref{eq7}, we arrive at the following result.

\begin{equation}\label{eq8}
ln H = \frac{3}{2}(w_0+1)ln(1+z)+\frac{3}{4}w_aln(1+z^2)+C
\end{equation}
where $C$ is the integration constant.\\
Now, we evaluate the expression at the present epoch, that is, by setting $H=H_0$ and $z=0$, where $H_0$ is the current value of $H$, we obtain $C=ln H_0$.
By substituting the value of the constant 
$C$ into \eqref{eq8}, we obtain the final expression, in which the Hubble parameter$(H)$ is explicitly defined as a function of redshift$(z)$ as:
\begin{equation}\label{eq9}
H = H_0 (1+z)^{\frac{3}{2}(w_0+1)}(1+z^2)^{\frac{3}{4}w_a},
\end{equation}
the deceleration parameter $q$ as:
\begin{equation}\label{eq10}
q = -1 + 1.5 \left[ \frac{w_a z (1 + z)}{1 + z^2} + (1 + w_0) \right],
\end{equation}.
\subsection{Logarithmic Parametrization}
The second parameterization  in this study is logarithmic. In this case the EoS parameter $w_{eff}$ of the effective fluid in terms of time $z$ is, 
\begin{equation}\label{eq12}
w_{\rm eff}=w_0-w_a ln\left(\frac{1}{1+z}\right),
\end{equation}
To the best of our knowledge, this parameterization was originally introduced by Efstathiou \cite{ref55}.
Now using \eqref{eq1} and \eqref{eq2} into the relation $w_{\rm eff}=p_{\rm eff}/\rho_{\rm eff} $ and comparing it with \eqref{eq12} we obtain
\begin{equation}\label{eq13}
-\frac{2\dot{H}+3H^2}{3H^2} = w_0-w_a ln\left(\frac{1}{1+z}\right),
\end{equation}
\begin{equation}\label{eq14}
-\frac{2\dot{H}}{3H^2}-1 = w_0-w_a ln\left(\frac{1}{1+z}\right)
\end{equation}
Now, we utilize \eqref{eq6} in the context of \eqref{eq14} to derive a more refined expression, as follows: 
\begin{equation}\label{eq15}
\frac{2}{3H}(1+z)\frac{dH}{dz}=w_0-w_a ln\left(\frac{1}{1+z}\right)+1
\end{equation}
By integrating both sides of \eqref{eq15}, the following relationship is established.
\begin{equation}\label{eq16}
ln H = \frac{3}{2}\left[(w_0+1)ln(1+z)+\frac{w_a}{2}(
ln(1+z))^2\right]+C,
\end{equation}
where $C$ is the integration constant, which is now checked at the current time using $H=H_0$ and $z=0$, with $H_0$ being the current value of $H$. This gives $C = \ln H_0$. Substituting this value of $C$ into equation \eqref{eq16} results in the final version of the equation, where the Hubble parameter $H$, and the deceleration are clearly shown as functions of the redshift $z$.

\begin{equation}\label{eq17}
H(z) = H_{0} \,(1+z)^{\tfrac{3}{2}(w_{0}+1) + \tfrac{3w_{a}}{4}\ln(1+z)},
\end{equation}

\begin{equation}\label{eq18}
q = -1 + 1.5 \left[ (1 + w_0) + w_a \log(1 + z) \right]
\end{equation},

\section{Datasets and Methodology}

In this section, we provide a general idea of the main steps in analyzing observational data. The steps are as follows: \\
Background Analysis: The overall likelihood at the background level is checked using the $\chi^2_{\text{total}}$ statistic, which combines information from several observational data sets. 
This can be expressed as follows:
 \begin{equation}\label{xi22}
 \chi^2_{\text{total}}(\mathbf{p}) = \chi^2_{\text{CC}}(\mathbf{p}) + \chi^2_{\text{DESI BAO}}(\mathbf{p}) + \chi^2_{\text{SN}}(\mathbf{p}) 
 \end{equation}
In this study, the set of free parameters was denoted by $\mathbf{p} = { H_{0}, w_{0}, w_{a} }$.  The subscripts CC, DESI BAO, and SN indicate which data come from cosmic chronometers, DESI Baryon Acoustic Oscillations, and SNe Ia, respectively. For the analysis, we employed 1088 background observational data points, comprising 1048 data points from SNe Ia, 7 points from DESI BAO observations, and 33 data points corresponding to H(z) measurements.\\

\textbf{ Statistical Tools:} 
The $\chi^2$ statistic is commonly used to assess the level of agreement between theoretical models and observational data. 
In addition, we use a method known as Markov Chain Monte Carlo (MCMC), which helps examine all possible values for the model's parameters.  This helps us to understand how uncertain we are about each parameter and how they might be connected. Together, these tools allow  us to work with the data in a clear manner and narrow down the best possible values for the model parameters based on how well they fit the data. In the next sections, we provide a quick look at the data sets used in this study.

\subsection{Type Ia Supernovae(SNIa) data}

The Type Ia Supernovae (SNIa) dataset plays a crucial role in studying the dynamic background of the universe and continues to provide valuable constraints for DE models. The SnIa dataset involves comparing the apparent magnitude with the absolute magnitude of the observed SnIa, which is known as the distance modulus and is theoretically given by:

\begin{equation}
\mu_{th}(z)=5\log_{10}{d_{L}(z)}+ \mu_{0}
\end{equation}
where $\mu_0 = 42.384 - 5 \log_{10} h$, and $d_L(z)$ is the luminosity distance, which is defined as follows:

\begin{equation}
d_{L}(z)=\frac{c}{H_{0}}(1+z)\int_{0}^{z}\frac{dz^{\prime}}{E(z^{\prime})}
\end{equation}

In this study, we employ the SNIa dataset of 1048 points sample~\cite{refc}.  The corresponding $\chi^2_{\text{SN}}$ can be derived as follows

\begin{equation}
\chi^{2}_{\mathrm{SN}}(\textbf{p})=\sum_{i,j}^{1048} \Delta\mu_i \left(C_{\text{stat + syst}}^{-1}\right)_{ij} \Delta\mu_j.
\end{equation}
Here, $\Delta \mu_i = \mu^{\text{th}}_i - \mu^{\text{obs}}_i$ represents the deviation between the theoretical and observed distance modulus values. The matrix $C_{\text{stat + syst}}^{-1}$ is the inverse of the covariance matrix corresponding to the Pantheon dataset, which accounts for statistical and systematic ( $C_{\text{stat + syst}} = C_\text{stat} + C_\text{syst}$) correlations between supernova measurements.\\

\subsection{\textbf{Cosmic Chronometer}}

In this study, we employ 33 observational data points of the Hubble parameter $H(z)$ listed in Table(\ref{tab:HubbleData}), covering the redshift range $0.07 \leqslant z \leqslant 1.965$. Given that these 
$H(z)$ measurements are mutually uncorrelated, the corresponding chi-squared statistic, $\chi^{2}_{CC}$, can be formulated as follows:

\begin{equation}
\chi^{2}_{CC}(\textbf{p}) = \sum_{i=1}^{33} \frac{[H_{th}(\textbf{p}, z_{i}) - H_{obs}(z_{i})]^{2}}{\sigma^{2}_{i}}
\end{equation}

Here, $H_{th}(\mathbf{p}, z_{i})$ represents  the  predicted values of the model at redshift $z_{i}$. The terms $H_{obs}(z_{i})$ and $\sigma_{i}$ represent the corresponding observed values and their associated Gaussian uncertainties, respectively, as listed in Table (\ref{tab:HubbleData}).

\begin{table}[hbt!]
\centering
\begin{tabular}{cccc}
\toprule
$z$ & $H(z)$ & $\sigma_H$ & References \\
\midrule
0.07 & 69 & 19.6 & \cite{ref56} \\
0.09 & 69 & 12 & \cite{ref57} \\
0.12 & 68.6 & 26.2 & \cite{ref56} \\
0.17 & 83 & 8 & \cite{ref57} \\
0.179 & 75 & 4 & \cite{ref58} \\
0.199 & 75 & 5 & \cite{ref58} \\
0.2 & 72.9 & 29.6 & \cite{ref56} \\
0.27 & 77 & 14 & \cite{ref57} \\
0.28 & 88.8 & 36.6 & \cite{ref56} \\
0.352 & 83 & 14 & \cite{ref58} \\
0.3802 & 83 & 13.5 & \cite{ref59} \\
0.4 & 95 & 17 & \cite{ref57} \\
0.4004 & 77 & 10.2 & \cite{ref59} \\
0.4247 & 87.1 & 11.2 & \cite{ref59} \\
0.4497 & 92.8 & 12.9 & \cite{ref59} \\
0.4700 & 89.0 & 49.6 & \cite{ref60} \\
0.4783 & 80.9 & 9 & \cite{ref59} \\
0.48 & 97 & 62 & \cite{ref61} \\
0.593 & 104 & 13 & \cite{ref58} \\
0.68 & 92 & 8 & \cite{ref58} \\
0.75 & 98.8 &33.6 & \cite{ref62} \\
0.781 & 105 & 12 & \cite{ref58} \\
0.8 & 113.1 & 28.5 & \cite{ref63} \\
0.875 & 125 & 17 & \cite{ref58} \\
0.88 & 90 & 40 & \cite{ref61} \\
0.9 & 117 & 23 & \cite{ref58} \\
1.037 & 154 & 20 & \cite{ref61} \\
1.3 & 168 & 17 & \cite{ref57} \\
1.363 & 160 & 33.6 & \cite{ref64} \\
1.43 & 177 & 18 & \cite{ref57} \\
1.53 & 140 & 14 & \cite{ref57} \\
1.75 & 202 & 40 & \cite{ref57} \\
1.965 & 186.5 & 50.4 & \cite{ref64} \\
\bottomrule
\end{tabular}
\caption{The H(z) data employed in this study (expressed in units of $\mathrm{km, s^{-1} Mpc^{-1}}$) are drawn in part from the compilation provided by \cite{ref59}.}
\label{tab:HubbleData}
\end{table}

\subsection{\textbf{DESI BAO}}

In this work, we employed baryon acoustic oscillation (BAO) measurements from the first data release (DR1) of the Dark Energy Spectroscopic Instrument (DESI) survey \cite{ref65}. DESI DR1 provides the most precise BAO measurements to date, obtained from over six million galaxies and quasar redshifts collected during the first year of observations. We examine at Baryon Acoustic Oscillation (BAO) results from the Dark Energy Spectroscopic Instrument (DESI) survey. This survey included observations of galaxies, quasars, and Lyman-$\alpha$ tracers. These observations, listed in Table I of Ref.~\cite{ref65}, are for both isotropic and anisotropic BAO measurements in the redshift range $0.1 < z < 4.2$. These were broken down into seven redshift bins. Table I of \cite{ref65}  summarizes these measurements, which include both isotropic and anisotropic BAO measurements separated into seven redshift bins. The isotropic BAO measurements are shown as $D_{\mathrm{V}}(z)/r_{\mathrm{d}}$, where $D_{\mathrm{V}}$ is the angle-averaged distance, set to the comoving sound horizon at the drag epoch.  $D_{\mathrm{M}}(z)/r_{\mathrm{d}}$ and $D_{\mathrm{H}}(z)/r_{\mathrm{d}}$ are  the anisotropic BAO measurements. $D_{\mathrm{M}}$ is the comoving angular diameter distance, and $D_{\mathrm{H}}$ is the Hubble horizon.  We also consider how $D_{\mathrm{M}}/r_{\mathrm{d}}$ and $D_{\mathrm{V}}/r_{\mathrm{d}}$ are related.  We call this dataset "DESI."
 We define the chi-squared function for each measurement in Table I of \cite{ref65} as follows: 
\begin{equation}
	\chi^2_{\text{DESI BAO}}(\textbf{p}) =  \Delta Q_i \left(C_{\text{DESI BAO}}^{-1}\right) \Delta {Q_i}^T.
\end{equation}

where,
\[
\Delta Q_i =
\begin{cases}
	\left( \frac{D_M}{r_d} \right)^{\mathrm{th}}(z_i) - \left( \frac{D_M}{r_d} \right)^{\mathrm{obs}}(z_i), & \text{for } D_M \text{ measurements}, \\[8pt]
	\left( \frac{D_H}{r_d} \right)^{\mathrm{th}}(z_i) - \left( \frac{D_H}{r_d} \right)^{\mathrm{obs}}(z_i), & \text{for } D_H \text{ measurements}, \\[8pt]
	\left( \frac{D_V}{r_d} \right)^{\mathrm{th}}(z_i) - \left( \frac{D_V}{r_d} \right)^{\mathrm{obs}}(z_i), & \text{for } D_V \text{ measurements}.
\end{cases}
\]

In our cosmological analysis, we efficiently explored the parameter space using the Markov Chain Monte Carlo (MCMC) method. Building on previous studies, we extended the analysis by incorporating a broader set of observational data and imposing more stringent priors on the model parameters. Specifically, we investigated the parameter set $(H_0, w_0, w_a)$ using the \textit{emcee} library \cite{ref66} for parallelized MCMC sampling. The priors for the sampled parameters are chosen as $w_0 \in [-3,1]$, $w_a \in [-3,2]$, and $H_0 \in [60,85]$ km/s/Mpc, To ensure the reliability and robustness of our results, we employ 80 walkers, each running for 100,000 steps. By jointly analyzing the Pantheon compilation of Type Ia supernovae, DESI BAO data, and cosmic chronometer (CC) measurements, we derive tighter constraints on the model parameters, thereby deepening our understanding of the universe's expansion history. We stress that the Markov Chain Monte Carlo method utilized in this study adheres to the accepted and popular methodology in the literature. The objective of the current analysis is to utilize a robust and established framework to extract reliable constraints and physical consequences of dark energy parametrizations using contemporary observational data, rather than to introduce a novel statistical or sampling technique.

\section{Result and Discussion}
In this section, we present  observational constraints on the two parameterizations considered in this study. We compare each case with two different sets of data: Pantheon+DESI BAO+CC and CC+DESI BAO. We know that because we looked at many different models and the results are quite similar, the discussion might be repetitive. However, it is still important to cover everything thoroughly. If you want to know more about specific models, you can find the detailed numbers, graphs showing how different factors relate to each other, and charts showing the possible values of the main parameters.

\begin{itemize}
   
    \item Table II and Fig. 2  present the results of the BA parameterization discussed in
section A
\item Table II and Fig. 4 summarize the results of the logarithmic parameterization 
detailed in Section B. 
  \end{itemize}

\begin{table}[htbp]
\centering
\renewcommand{\arraystretch}{1.4}
\begin{tabular}{|c|c|c|}
\hline
\textbf{Data} & \textbf{Cosmic Chronometer(CC)} & \textbf{CC + Pantheon + DESI BAO} \\
\hline
\textbf{Parameterization} & \begin{tabular}{c} BA parameterization \\ \textcolor{magenta}{Logarithmic parameterization} \end{tabular} & \begin{tabular}{c} BA parameterization \\ \textcolor{magenta}{Logarithmic parameterization} \end{tabular} \\
\hline

$H_0$ [km/s/Mpc] & $69.01^{+0.0097}_{-0.011}$ & $72.22^{+0.0097}_{-0.011}$ \\
                 & \textcolor{magenta}{$67.078^{+0.0097}_{-0.011}$} & \textcolor{magenta}{$72.27^{+0.0097}_{-0.011}$} \\
\hline
$w_o$ &$-0.7091 \pm 0.0079$  & $-0.7667 \pm 0.0079$ \\
           & \textcolor{magenta}{$-0.6515 \pm 0.0079$} & \textcolor{magenta}{$-0.7911 \pm 0.0079$} \\
\hline
$w_a$ & $0.5212 \pm 0.0095$ & $0.6761 \pm 0.0095$ \\
      & \textcolor{magenta}{$0.6267 \pm 0.0095$} & \textcolor{magenta}{$1.0354 \pm 0.0095$} \\

\hline
\end{tabular}
\caption{An overview of the MCMC results obtained
from the CC and CC+Pantheon+DESI BAO datasets.}
\label{tab:ft_vs_lcdm}
\end{table}

\subsection{Result for BA parameterization}

\begin{figure}[hbt!]
    
  (a)  \includegraphics[width=0.5\linewidth]{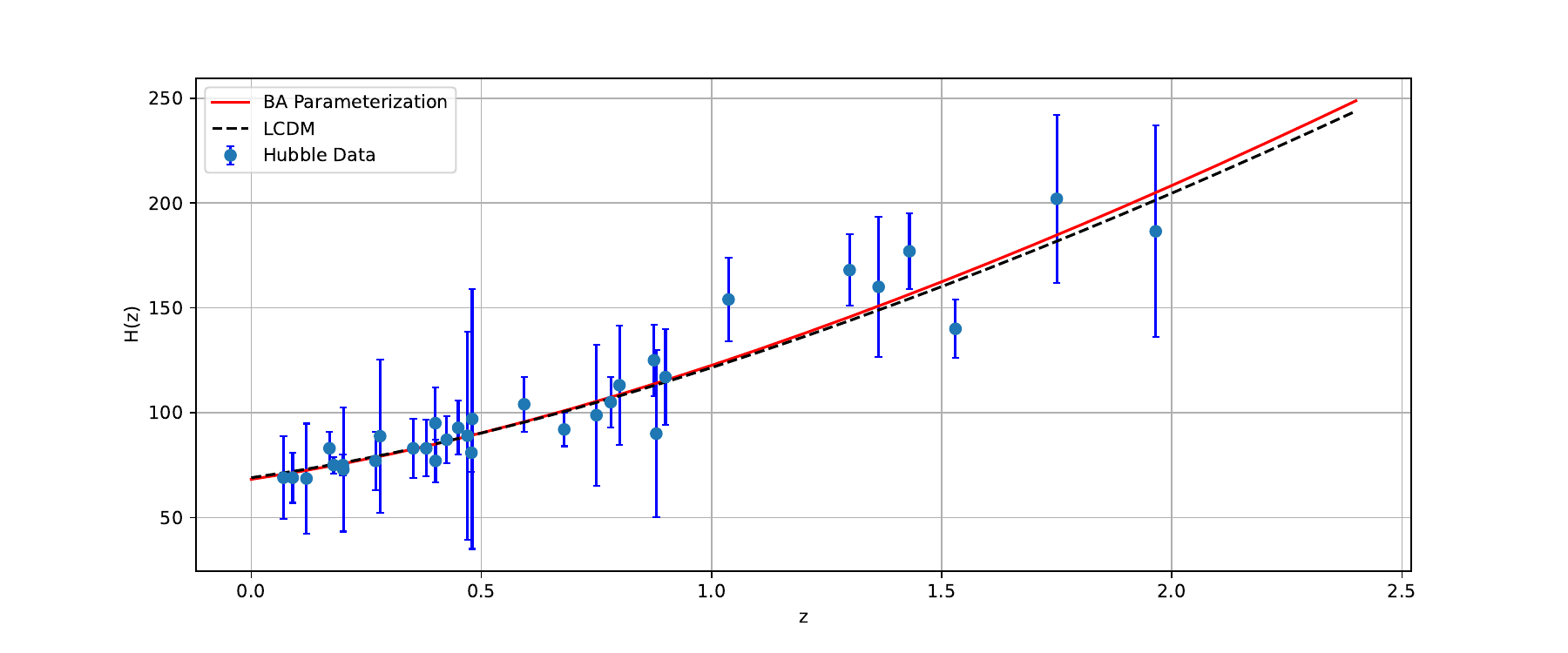}\\ 
   (b)  \includegraphics[width=0.5\linewidth]{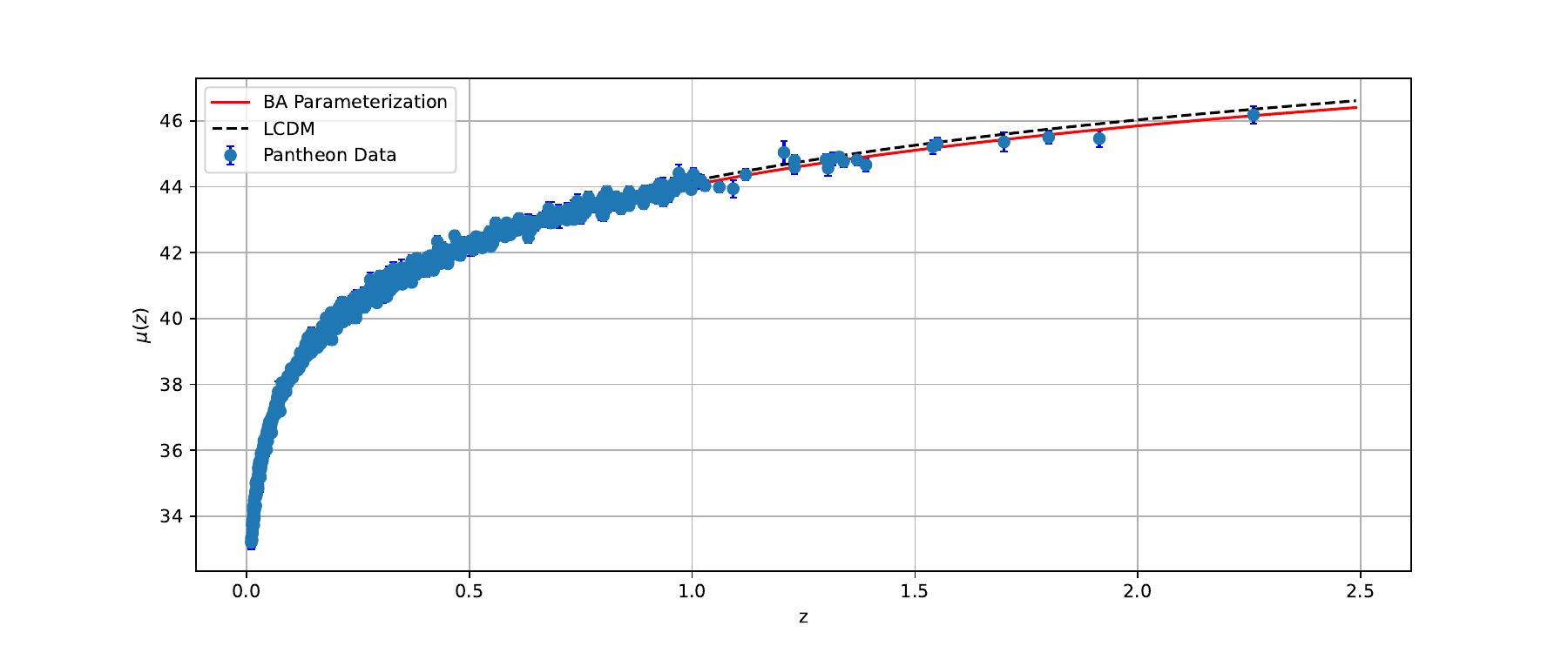}
    \caption{The plot of Hubble parameter $H(z)$, and $\mu(z)$ versus $z$ for BA parameterization}
    \label{fig:enter-label}
\end{figure}

The values of the Hubble parameter $H(z)$ observed at various redshift points $z$, as listed in Table 1 (Hubble Data), were used to construct the corresponding $H(z)$ curve. Figure 1(a) presents this comparison, where the vertical ticks represent the observed $H(z)$ data, and the solid line depicts the theoretical $H(z)$ calculated from the model, plotted against the redshift $z$. BA parameterization involves three independent parameters: $w_0$, $H_0$, and $w_a$. To constrain these parameters, we employ the Markov Chain Monte Carlo (MCMC) technique, specifically using the \textit{emcee} Python package developed by Mackey et al. \cite{ref66}. By running the MCMC analysis, we obtained  best-fit values at the $1\sigma$ confidence level. Using the CC datasets alone, the results are $w_0 = -0.7091 \pm 0.0079$, $H_0 = 69.01^{+0.0097}_{-0.011}$, and $w_a = 0.5212 \pm 0.0095$. When combining the CC, Pantheon, and DESI BAO datasets, the best-fit parameters were $w_0 = -0.7667 \pm 0.0079$, $H_0 = 72.22^{+0.0097}_{-0.011}$, and $w_a = 0.6761 \pm 0.0095$. For these combined datasets, we also present two-dimensional likelihood contours corresponding to the $1\sigma$ (68$\%$) and $2\sigma$ (95$\%$) confidence intervals. Figure 2 illustrates the resulting best-fit plots based on  observational data.

\begin{figure}[hbt!]
    \includegraphics[width=0.5\linewidth]{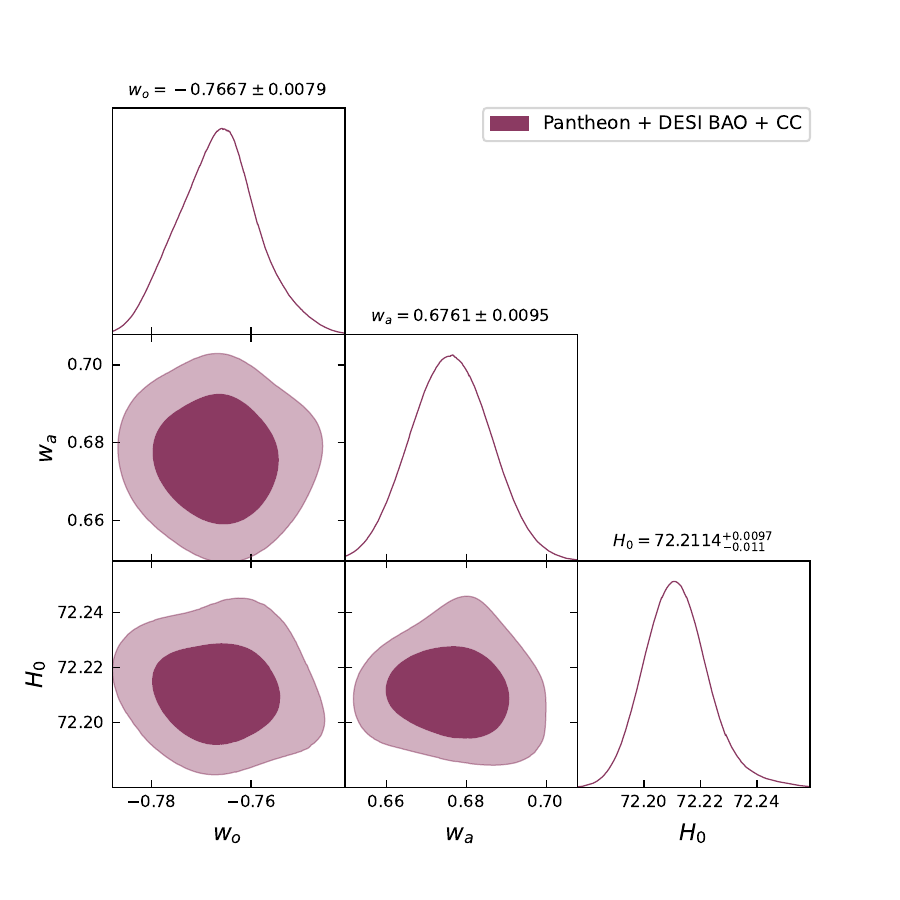}
    \caption{The best fit plots for 1-D marginalized distribution, and 2-D contours with $68\%$ CL and $95\%$ CL for the model parameters with the combined CC + DESI BAO + Pantheon datasets.}
    \label{fig:enter-label}
\end{figure}
\subsection{Result for Logarithmic parameterization}
\begin{figure}[hbt!]
    
    \includegraphics[width=0.5\linewidth]{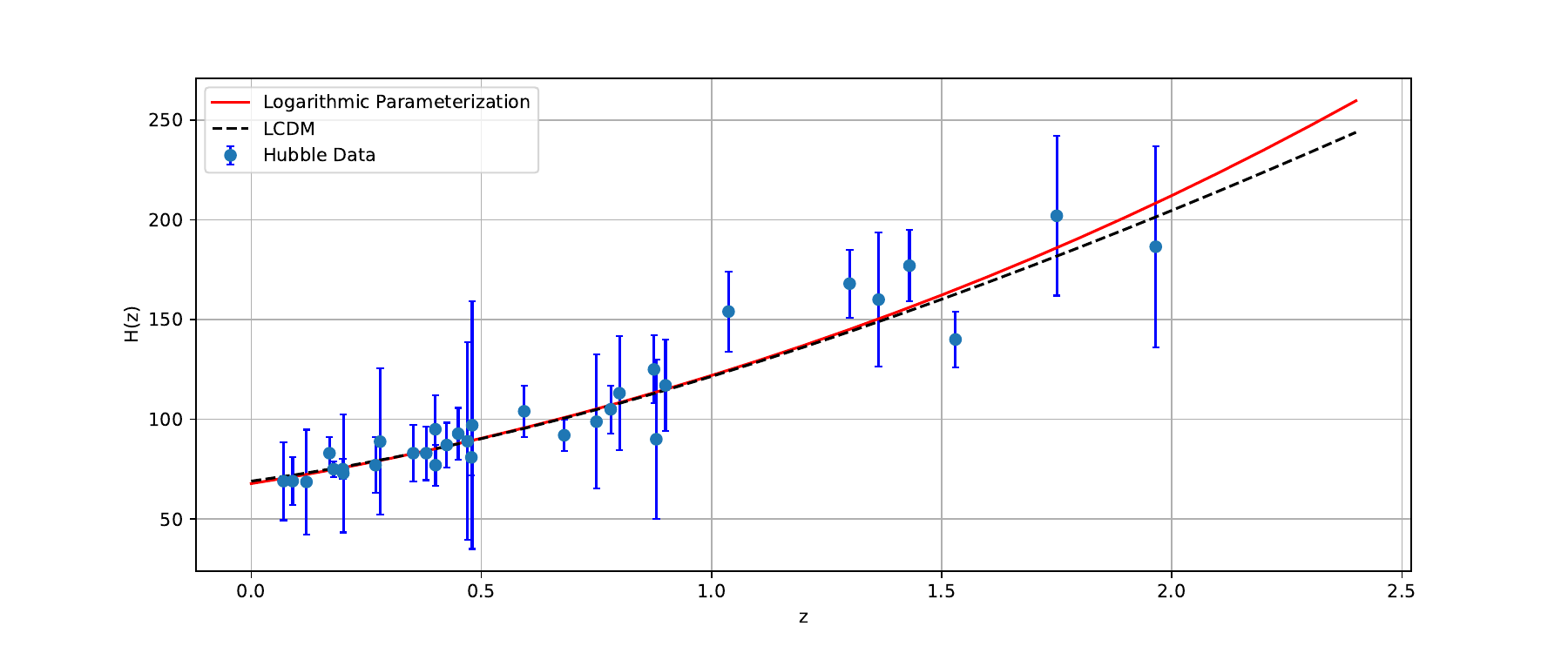}
    \includegraphics[width=0.5\linewidth]{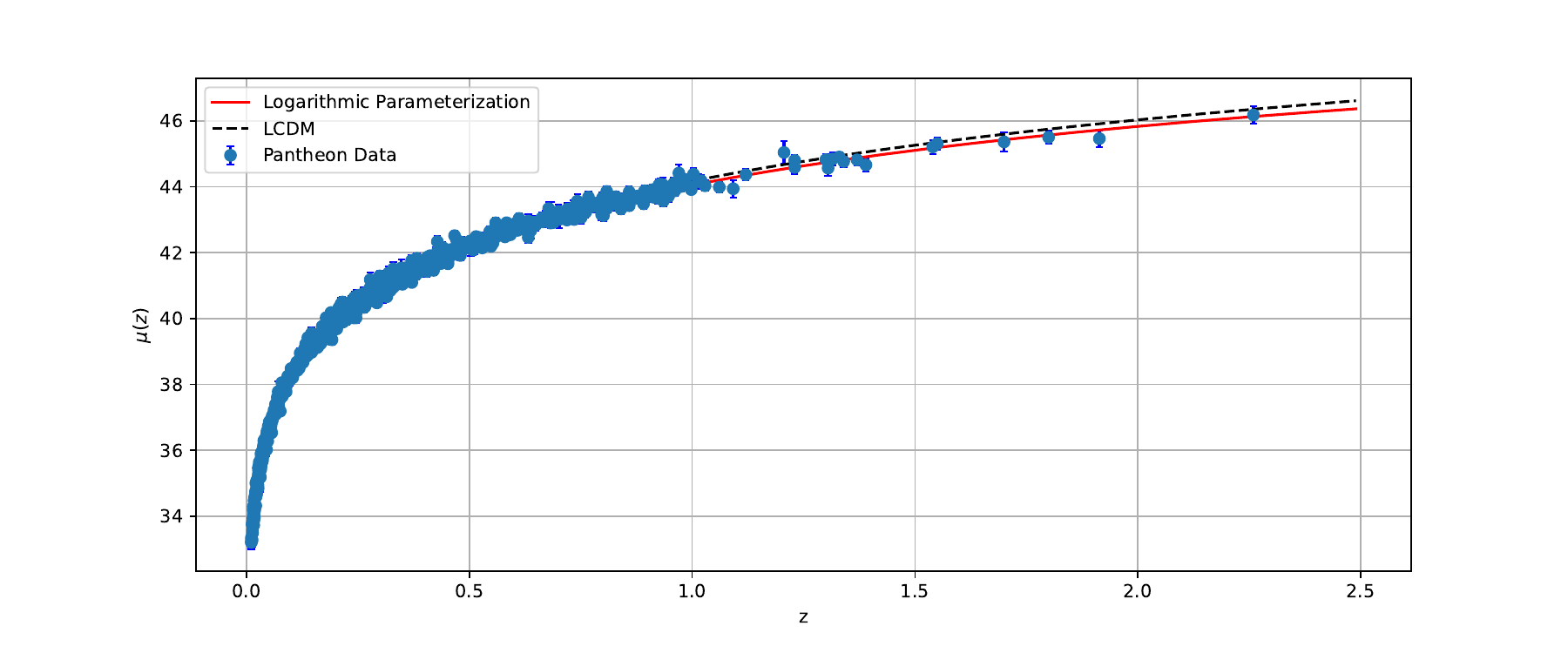}
    \caption{The plot of Hubble parameter $H(z)$, and $\mu(z)$ versus z for Logarithmic parameterization}
    \label{fig:enter-label}
\end{figure}

\begin{figure}[hbt!]
    \centering
    \includegraphics[width=0.5\linewidth]{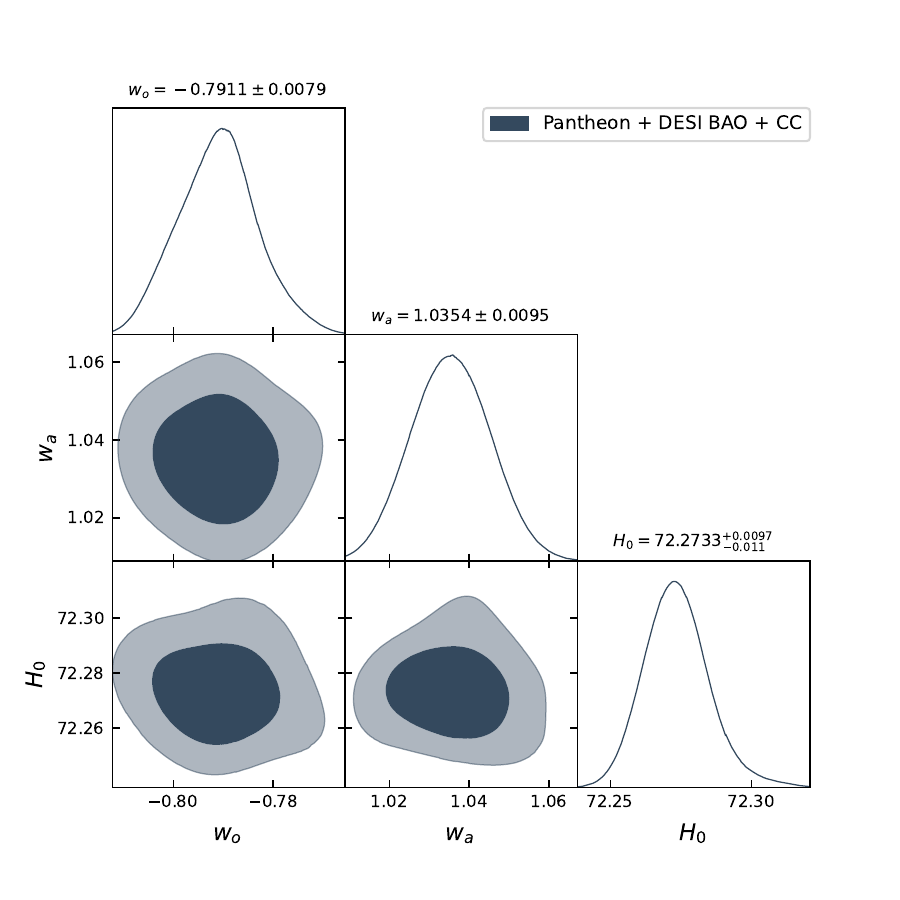}
    \caption{The best fit plots for 1-D marginalized distribution, and 2-D contours with $68\%$ CL and $95\%$ CL for the model parameters with the combined CC + DESI BAO + Pantheon datasets.}
    \label{fig:enter-label}
\end{figure}

The observed value of the Hubble parameter $H(z)$ at different redshift parameters $z$ given in Table 1, was employed to draw the curve corresponding to the curve of $H(z)$. Figure 3(a) shows the plot of observed $H(z)$ (vertical small lines) and the calculated $H(z)$ (solid curve) versus the redshift parameter $z$. Logarithmic parameterization involves three free parameters: $w_0$, $H_0$, and $w_a$. To constrain these parameters and determine their best-fit values, we applied the Markov Chain Monte Carlo (MCMC) technique. Specifically, we use the ensemble sampler implemented in the \textit{emcee} Python library, developed by  Mackey et al. \cite{ref66}. Using MCMC, we obtain the following best-fit values with $1\sigma$ confidence intervals: for the CC datasets, $w_0 = -0.6515 \pm 0.0079$, $H_0 = 67.08^{+0.0097}_{-0.011}$, and $w_a = 0.6267 \pm 0.0095$; and for the combined CC + Pantheon + DESI BAO datasets, $w_0 = -0.7911 \pm 0.0079$, $H_0 = 72.27^{+0.0097}_{-0.011}$, and $w_a = 1.0354 \pm 0.0095$. Additionally, we present the two-dimensional likelihood contours for these combined datasets, showing $1\sigma$ and $2\sigma$ confidence regions corresponding to confidence levels of 68$\% $and 95$\%$. Figure 4 illustrates the resulting best-fit curves derived from the observational data.\\

The constraints obtained on the dark energy equation of state parameters, $w_0$ and $w_a$, provide insight into the role of dynamical dark energy in alleviating the Hubble tension. In particular, our analysis indicates a mild preference for $w_a \neq 0$, suggesting that the dark energy equation of state evolves with redshift rather than remaining constant as in the $\Lambda$CDM model ($w = -1$). This evolution contributes to our best-fit values of the Hubble constant obtained from the DE models are $H_0 = 72.2 2^{0.0097}_{0.011}$ km/s/Mpc for BA parameterization and $H_0 = 72.2 7^{0.0097}_{0.011}$ for Logarithmic which align more closely with local measurements with the Planck CMB $\Lambda$CDM value ($H_0 = 67.4 \pm 0.5$ km/s/Mpc). This indicates that the dynamical dark energy models considered in this study partially alleviate the Hubble tension, reducing it from $4.4\sigma$ (Planck vs. SH0ES) to $1.8\sigma$. These results suggest that late-time modifications of the dark energy sector can play an important role in reconciling the discrepancy between the early and late-universe measurements of $H_0$.\\

The numerical constraints obtained in this section serve as the basis for extracting the physical implications of the dark energy dynamics. In particular, the best-fit parameters are used in the subsequent sections to reconstruct the redshift evolution of the equation of state and  deceleration parameter, and to perform the statefinder diagnostics. These analyses allow us to infer the nature of the dark energy (quintessence or phantom behavior), the transition from deceleration to acceleration, and the future cosmic fate predicted by the models, which are discussed in detail in Sections V and VI.

\section{Deceleration and EOS Parameter}
Early in the 20th century, Edwin Hubble introduced the deceleration parameter, represented by the symbol q. To understand the mechanics of the expansion of the universe, it is a crucial cosmological quantity.  The mathematical expression for this equation is $q = -1- H/ \dot{H}^2$.  Understanding this characteristic is essential for comprehending the cosmic history as well as its future development. The deceleration parameter $q$ yields information about the dynamics of the universe expansion.  When the gravitational pull of matter was thought to be the main driver of cosmic expansion, a positive $q$ meant that the expansion slowed.  A deceleration value of zero, on the other hand, indicates a "critical Universe," which is defined by a constant rate of expansion that is neither accelerating nor decelerating.  In contrast, a negative $q$ denotes an accelerating expansion. This idea gained popularity after dark energy was discovered in the late 20th century, providing an explanation for the apparent acceleration of cosmic expansion.\cite{ref68}\\

The behavior of the deceleration parameter with respect to redshift for both the parameterizations, is shown in Fig. 5. The transition redshift is represented by $z_{t}$, which marks the shift from the decelerating to accelerating phase. Additionally, both parameterizations exhibit a de Sitter phase, defined by a deceleration parameter of $q$.
Using Pantheon + DESI BAO + CC, the value of $q_0$ is $-0.31$ for both parameterizations, indicating a shift to an accelerating universe. These measurements also strengthen the hypothesis that the universe switches from deceleration to acceleration at $z = 1$ \cite{ref69,ref70}. The present value of $q_0$ is negative, indicating the presence of dark energy.\\
Fig. 5 compares the evolution of the redshift of the effective equation of state for the BA and Logarithmic parameterization using constraints from the Pantheon + DESI BAO + CC datasets. Both parameterizations predict a behavior close to the cosmological constant at the present time $(z=0)$, but diverge significantly at future redshifts. Notably, the Logarithmic parameterization enters the phantom regime $w_{eff}$ for $z<0$, implying a possible future dominated by the phantom dark energy. In contrast, the BA parameterization remained closer to the quintessence regime throughout, indicating a more stable dynamic evolution of dark energy.

\begin{figure*}[hbt!]
    
   (a) \includegraphics[width=0.46\linewidth]{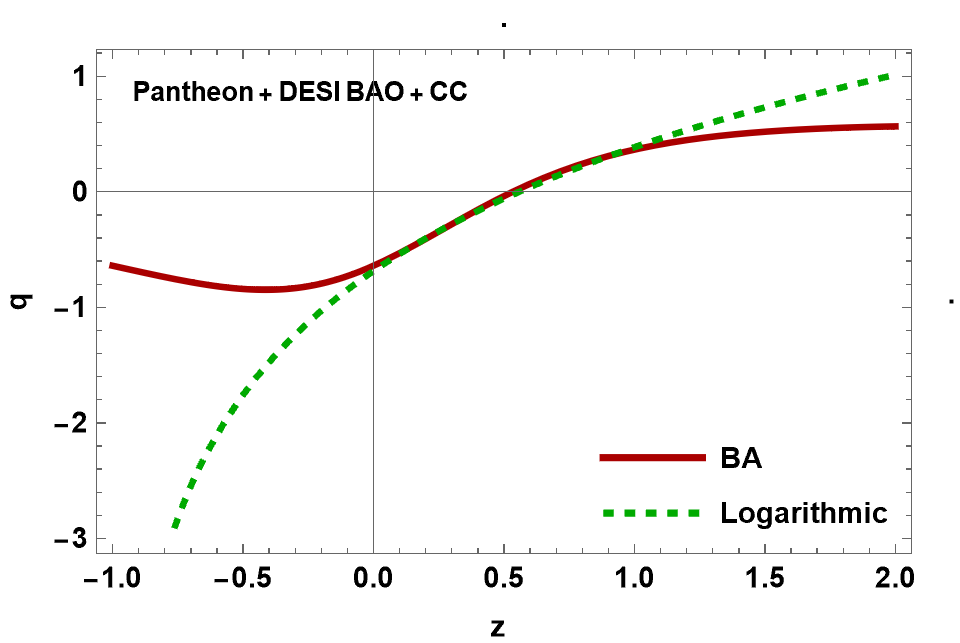}
   (b)\includegraphics[width=0.46\linewidth]{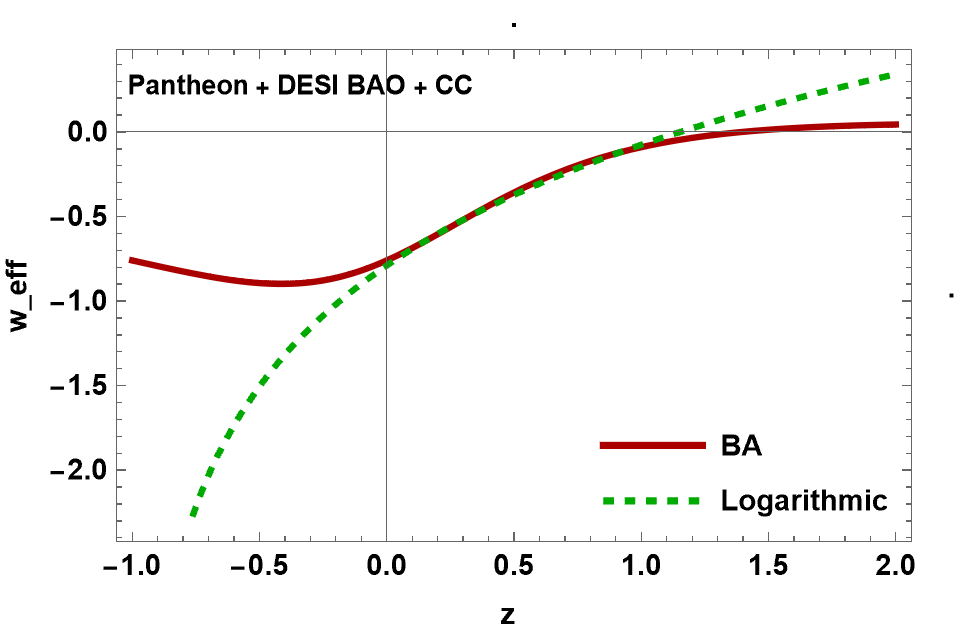}
    
    \caption{The plot displays the evolution of the effective Deceleration $(q)$ and EoS $(w_{eff})$ parameters with respect to the redshift $z$ using the values constrained from the Pantheon+DESI BAO+CC datasets.}
    
\end{figure*}

\section{State-finders}
\begin{figure*}[hbt!]

  (a)  \includegraphics[width=0.46\linewidth]{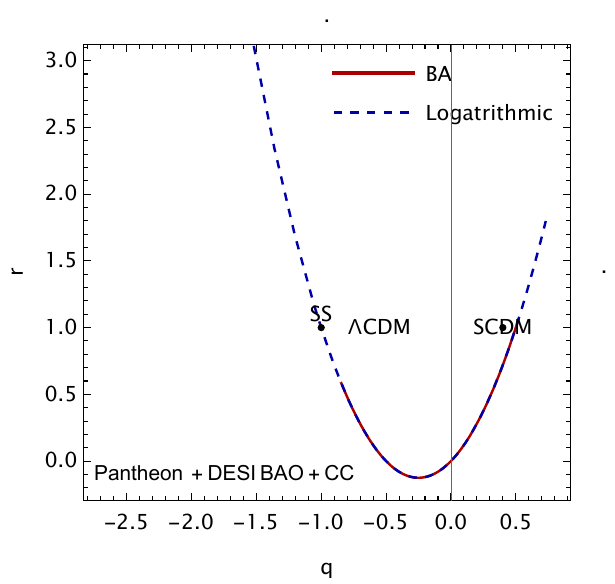}
   (b) \includegraphics[width=0.46\linewidth]{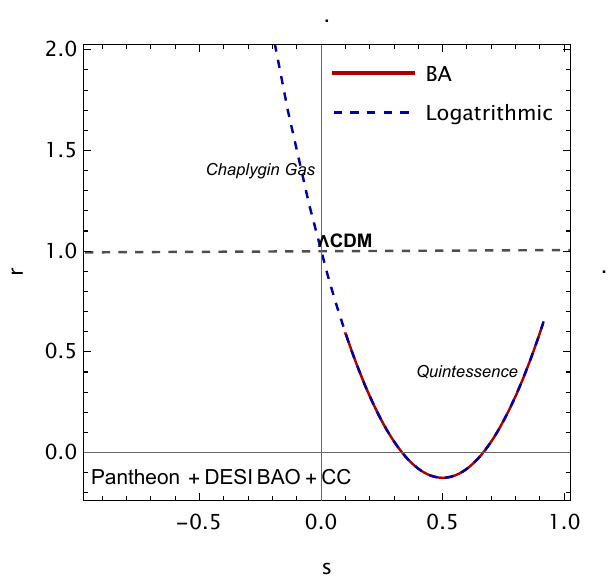}
    
    \caption{The plot displays the evolution of the $r–s$ and $r-q$ plane using the values constrained from the Pantheon+DESI BAO+CC datasets with redshift $z$.}
    
\end{figure*}

The statefinder parameters $\{r,s\}$, first introduced in \cite{ref71,ref72}, provide a geometric diagnostic to distinguish between dark energy models. Statefinders are diagnostic tools used in cosmology to analyze the behavior of cosmological parameters to examine the evolution of the universe. They describe the expansion of the cosmos geometrically, especially with regard to the dark energy equation of state, which is crucial to the universe's current energy makeup. In contrast to more conventional parameters such as the equation of state, $w =\frac{p}{\rho},$ of dark energy, the statefinder parameters provide a more thorough explanation of cosmic expansion.  Cosmologists can  test several cosmological theories (including $\Lambda$CDM, quintessence, and other dark energy models) and forecast the future evolution of the universe by tracking the evolution of these parameters. Finally, statefinders offer a productive way to investigate cosmic expansion and to distinguish between various theoretical models of cosmic dynamics and dark energy. It is defined as 
\begin{equation}
	r(z)= 1 - 2(1+z)\frac{H'(z)}{H(z)}
      + \left[ \frac{H''(z)}{H(z)} + \left( \frac{H'(z)}{H(z)} \right)^2 \right](1+z)^2 .
\end{equation}
\begin{equation}
	s(z)=\frac{r(z)-1}{3(q(z)-1/2)} 
\end{equation}

The vertical line in Fig.6 separates the $r-s$ trajectories into two sections.  In the $r-s$ plane, the region $r>1, s<0$ exhibits behavior associated with a Chaplygin gas (CG) model \cite{ref73}, while the region $r<1, s>0$ exhibits behavior associated with the quintessence model. Furthermore, the evolution trajectory was plotted on the $r$–$q$ plane. The $r$–$q$ trajectories in Fig.6 are separated into two sections by  point $(r, q) = (1, -1)$. In the $r$–$q$ plane, the regions $r > 1$, $q < -1$ exhibit behavior resembling the phantom model, whereas the region $r < 1$, $q > -1$ exhibits behavior resembling quintessence\cite{ref71,ref72}.  Statefinder analysis provides two important insights into the behavior of DE models. First, the trajectories in the $(r, s)$ plane show that the Barboza–Alcaniz (BA) and logarithmic models are geometrically distinct, despite both fitting the current observational data. This distinction reflects the differences in the underlying dynamics of the dark energy in these models. Second, the Logarithmic model exhibits a phantom-like behavior ($w < -1$) in the future ($z < 0$), suggesting a potential Big Rip scenario where the scale factor diverges in a finite time. By contrast, the BA model remains within the quintessence regime ($w > -1$) and asymptotically approaches a de Sitter-like expansion. Compared with the $\Lambda$CDM fixed point at $(r = 1, s = 0)$, both models show significant deviations, highlighting that dynamical DE models can lead to markedly different cosmic evolution in the future. These results emphasize that phantom/quintessence divergence has important implications for the ultimate fate of the universe, motivating further studies of late-time DE dynamics. Despite being  derived only from the backdrop expansion, the statefinder parameters offer important physical information about the effective nature of the dark energy fluid. Specifically, the trajectories in the $r–s$ and $r–q$ planes show whether the underlying fluid behaves as  phantom-like fluid,  quintessence-type fluid, or  cosmological constant-like component. While the asymptotic trends of the trajectories store information about the future cosmic fate, such as an approach to a de Sitter phase or a phantom-dominated expansion, the departure from the $\Lambda$CDM fixed point shows dynamic dark energy behavior. Therefore, by offering a physically significant description of the dark energy fluid at an effective level, the statefinder analysis supplements the observable limitations.\\

\textbf{Statistical Analysis}: In the final step of the observational analysis, we compared the BA, Logarithmic and CPL parameterizations using standard information criteria such as the  Akaike Information Criterion (AIC) \cite{add1} and the Bayesian, or Schwarz, Information Criterion (BIC) \cite{add2}. Both are widely employed in statistics and data analyses as described below.
$$\text{AIC} = -2 \ln \mathcal{L}+ 2 d = \chi^2_{min} + 2 d,$$

$$\text{BIC} = -2 \ln \mathcal{L}+ d \ln N = \chi^2_{min} + d \ln N,$$

In this expression, $\mathcal{L} = \exp\left(-\chi_{\text{min}}^2/2\right)$ yields the maximum likelihood, where $d$ is the number of free parameters in the model and $N$ is the total number of data points used in the experiment.  Using the typical $\Lambda$CDM cosmology as a baseline scenario is crucial for assessing and contrasting other hypotheses. The relative performance of any model $M$ can be evaluated by computing the difference $\Delta X = X_M - X_{\Lambda\text{CDM}}$, where $X$ is the AIC or BIC criterion. Based on the value of $\Delta X$, the model's level of support can be interpreted as follows: For the AIC, strong support if $\Delta \mathrm{AIC} \leq 2$ (indicating a good fit), moderate support if $4 \leq \Delta \mathrm{AIC} \leq 7$, and essentially no support if $\Delta \mathrm{AIC} \geq 10$. For BIC, the evidence is considered positive if \( 2 \leq \Delta \mathrm{BIC} \leq 6 \), strong if \( 6 < \Delta \mathrm{BIC} \leq 10 \), and \emph{very strong} if \( \Delta \mathrm{BIC} > 10 \). \\

\begin{table*}[hbt!]
	\caption{The difference, $\Delta \text{AIC} = \text{AIC}_{\text{Model}} - \text{AIC}_{CPL}$ and 
	$\Delta \text{BIC} = \text{BIC}_{\text{Model}} - \text{BIC}_{CPL}$ for different cosmological models with respect to CPL parameterization from all considered data sets.}
	\label{tab3}
	\scalebox{0.85}{
		\begin{tabular}{lcc}
			\hline
			\toprule
			\textbf{Dataset } 
			& \textbf{CC} 
			& \textbf{CC+Pantheon+DESI BAO} \\
			\hline
			
			\textbf{Model} 
			& \textbf{BA parameterization} 
			& \textbf{BA parameterization} \\
			
			& \textcolor{teal}{\textbf{Logarithmic parameterization}} 
			& \textcolor{teal}{\textbf{Logarithmic parameterization}} \\
			
			& \textcolor{magenta}{\textbf{CPL parameterization}} 
			& \textcolor{magenta}{\textbf{CPL parameterization}} \\
			\hline
			
			\vspace{0.1cm}
			{\boldmath$\rm AIC$}
			& $22.31$ 
			& $1061.52$ \\
			
			& \textcolor{teal}{$21.02$} 
			& \textcolor{teal}{$1060.96$} \\
			
			& \textcolor{magenta}{$20.5$} 
			& \textcolor{magenta}{$1059.75$} \\
			\hline
			
			\vspace{0.1cm}
			{{\boldmath$\rm \Delta AIC$}}
			& $1.81$ 
			& $1.77$ \\
			
			& \textcolor{teal}{$0.52$} 
			& \textcolor{teal}{$1.21$} \\
			
			& \textcolor{magenta}{$0$} 
			& \textcolor{magenta}{$0$} \\
			\hline
			
			\vspace{0.1cm}
			{{\boldmath$\rm BIC$}}
			& $26.8$ 
			& $1076.38$ \\
			
			& \textcolor{teal}{$25.50$} 
			& \textcolor{teal}{$1075.82$} \\
			
			& \textcolor{magenta}{$24.98$} 
			& \textcolor{magenta}{$1074.36$} \\
			\hline
			
			\vspace{0.1cm}
			{{\boldmath$\rm \Delta BIC$}}
			& $1.82$ 
			& $2.02$ \\
			
			& \textcolor{teal}{$0.52$} 
			& \textcolor{teal}{$1.46$} \\
			
			& \textcolor{magenta}{$0$} 
			& \textcolor{magenta}{$0$} \\
			\hline
			\hline
			
			\vspace{0.1cm}
			{{\boldmath$\chi^2_{\rm min}$}}
			& $16.31$ 
			& $ 1055.52$ \\
			
			& \textcolor{teal}{$15.02$} 
			& \textcolor{teal}{$1054.96$} \\
			
			& \textcolor{magenta}{$14.5 $} 
			& \textcolor{magenta}{$1053.75$} \\
			\hline
			\hline
		\end{tabular}
	}
\end{table*} 
To statistically compare different dark energy parameterizations, we used the AIC and BIC, which balance the goodness of fit against model complexity. The corresponding values of AIC, BIC, and the minimum chi-square are summarized in Table \ref{tab3} for the CC and CC+Pantheon+DESI BAO datasets. It is evident that the BA and Logarithmic parameterizations yield slightly higher values of $\chi^2_{\rm min}$ than the CPL parameterization. Consequently, their $\Delta$AIC and $\Delta$BIC values remain positive with respect to CPL. Although the differences are small, indicating statistical competitiveness, they do not provide evidence for a preference of the extended models over  CPL parameterization.

\section{Conclusion}
In this study, we performed a detailed observational analysis of two prominent dark energy parameterizations the Barboza-Alcaniz (BA) and logarithmic models within a spatially flat Friedmann-Lemaître-Robertson-Walker (FLRW) cosmology. The primary aim was to evaluate the performance and constraining power of these models using the latest background observational data, including 1048 SNe Ia from the Pantheon sample, 7 DESI BAO points,
and 33 Hubble parameter data from cosmic chronometers. Our analysis followed a rigorous methodology involving the $\chi^2$ minimization technique and MCMC sampling to derive constraints on the model parameters $\omega_{0}$, $\omega_{a}$ $H_{0}$. We obtained the best-fit parameters with high statistical confidence for both models across individual and combined datasets. Notably:

\begin{itemize}
    \item For the \textbf{Barboza-Alcaniz (BA)} model, the combined dataset ( CC + Pantheon + DESI BAO) analysis yielded the following results:
    \[
    w_0 = -0.7667 \pm 0.0079,\quad H_0 = 72.22^{+0.0097}_{-0.011},\quad w_a = 0.6761 \pm 0.0095.
    \]

    \item For the \textbf{Logarithmic} model, the combined dataset ( CC + Pantheon + DESI BAO) analysis yielded
    \[
    w_0 = -0.7911 \pm 0.0079,\quad H_0 = 72.27^{+0.0097}_{-0.011},\quad w_a = 1.0354 \pm 0.0095.
    \]
\end{itemize}

The results highlight that both models are compatible with current cosmological observations, but the logarithmic parameterization provides slightly tighter constraints on  DE evolution, particularly because of its stronger sensitivity in the late-time universe. Additionally, through the \textit{statefinder diagnostic}, we distinguish the geometrical trajectories of each DE model in the $(r, s)$ and $(r, q)$ planes, revealing subtle but significant differences. Statefinder analysis supports the robustness of these models and provide evidence that future precision data could further break the degeneracy between them.
Our findings indicate that both parameterizations are in good agreement with current observational data and successfully capture the transition from deceleration to acceleration in the cosmic expansion history.  The present value of the deceleration parameter $q_0$ is negative $(q_{0} \approx -0.31)$, consistent with an accelerating universe dominated by dark energy. The estimated redshift of the transition $z_{t}\approx 1$ 
further corroborates the results of previous studies. A detailed comparative analysis revealed that while the BA parameterization maintains $w_{eff}$ close to the quintessence regime, leading to a more stable late-time evolution, the Logarithmic parameterization exhibits a future evolution into the phantom regime ($w_{eff}<-1$), implying a more dynamically evolving dark energy component. Notably, the logarithmic parameterization yields slightly tighter constraints on $w_{a}$, suggesting a greater sensitivity to the evolution of dark energy.\\

Additionally, the statefinder diagnostic was employed to examine the geometrical properties of both models, highlighting subtle differences in their dynamic trajectories and providing further means of distinguishing between them. Overall, this study demonstrates that both parameterizations remain compatible with current cosmological observations, yet exhibit distinct implications for the future evolution of the universe. The results not only enhance our understanding of the possible behaviors of dark energy but also provide valuable insights for future observational strategies aimed at discriminating between competing dark energy scenarios.
Although the functional forms of the BA and Logarithmic model were proposed in earlier studies, this study demonstrates that their unified confrontation with modern high-precision data and geometric diagnostics leads to qualitatively new physical interpretations of late-time cosmic evolution, thereby establishing the originality of the present analysis beyond mere parameter estimation. Our analysis shows different physical implications for the cosmic fate indicated by the two theories, going beyond the parameter estimation. Logarithmic parameterization forecasts a future phantom regime that could result in a Big Rip scenario, whereas the BA parameterization asymptotically approaches a de Sitter phase. These qualitative differences highlight that phenomenological parameterizations, when combined with modern data and geometric diagnostics, can yield valuable physical insights into the nature of dark energy.\\
From a statistical model-comparison perspective, the Akaike and Bayesian information criteria indicate that both the Barboza–Alcaniz and logarithmic parameterizations are statistically competitive with the CPL parameterization, but they are not statistically preferred over it. This result suggests that, while extended phenomenological parameterizations remain viable descriptions of dark energy, current data do not provide strong statistical evidence to favor them over simpler benchmark models. Nevertheless, the distinct late-time dynamical behaviors revealed through  equation of state evolution and statefinder diagnostics provide complementary physical insights that go beyond parameter estimation alone.
The novelty of this study lies in: the use of the latest DESI BAO data to reassess two widely used dark energy parameterizations,  a direct physical and geometric comparison between the models within a unified framework, and  the extraction of distinct late-time and future cosmological behaviors through statefinder diagnostics.

\section*{Declaration of competing interest}
The authors declare that they have no known competing financial
interests or personal relationships that could have influenced the work reported in this study.

\section*{Data availability}
We employed publicly available Pantheon data, Cosmic Chronometer (CC) data, and DESI BAO data presented in this study. The CC data are compiled from publicly available cosmic chronometer measurements in the literature, with a representative compilation accessible at: https://github.com/AhmadMehrabi Cosmic chronometer data. The Pantheon compilation (distance moduli and covariance matrices),  which is publicly available on GitHub: https://github.com/brinckmann/montepython$\_$public/tree/3.6/montepython/likelihoods/Pantheon and the DESI BAO data which is publicly available on Github: https://github.com/LauraHerold/MontePython$\_$desilike/blob/main/likelihood/bao$\_$desi$\_$all/ . No additional datasets are used in this study.

\section*{acknowledgments}
The authors (AD \& AP) are thankful to Inter-University Centre for Astronomy and Astrophysics (IUCAA), Pune, India, for providing support and facilities under the Visiting Associateship program. The author (S. Verma) was supported by a Senior Research Fellowship (UGC Ref No. 192180404148) from the University Grants Commission, Govt. of India. The authors are appreciative of the anonymous reviewer's insightful criticism and recommendations, which enhanced our work in its current form.

\end{document}